\documentclass[10pt]{article}
\usepackage[letterpaper]{geometry}
\usepackage{hicss51}
\usepackage{times}
\usepackage[none]{hyphenat}
\usepackage{url}
\usepackage{latexsym}
\usepackage{indentfirst}
\usepackage{graphicx}
\graphicspath{{images/}}

\usepackage{tabularx}
    
\usepackage{hyperref}
\usepackage{amsmath}
\usepackage{makecell}
\usepackage{graphicx}
\usepackage{comment}
\usepackage[belowskip=-10pt,aboveskip=0pt]{caption}

\title{VULNERLIZER: Cross-analysis Between Vulnerabilities and Software Libraries}

\author{Irdin Pekaric \\
  Department of Computer Science \\ University of Innsbruck \\
  {irdin.pekaric@uibk.ac.at} \\\And
  Philipp Steinm\"uller \\
  Department of Computer Science \\ University of Innsbruck \\
  { philipp.steinmueller@student.uibk.ac.at }\\\And 
  Michael Felderer \\
 Department of Computer Science \\ University of Innsbruck \\
  {michael.felderer@uibk.ac.at} \\}

\date{}

\begin{document}
\maketitle
\begin{abstract}
	
The identification of vulnerabilities is a continuous challenge in software projects. This is due to the evolution of methods that attackers employ as well as the constant updates to the software, which reveal additional issues. As a result, new and innovative approaches for the identification of vulnerable software are needed. In this paper, we present VULNERLIZER, which is a novel framework for cross-analysis between vulnerabilities and software libraries. It uses CVE (Common Vulnerabilities and Exposures) and software library data together with clustering algorithms to generate links between vulnerabilities and libraries. In addition, the training of the model is conducted in order to reevaluate the generated associations. This is achieved by updating the assigned weights. Finally, the approach is evaluated by making the predictions using the CVE data from the test set. The results show a great potential of the approach in predicting future vulnerable libraries based on an initial input CVE entry or a software library. The trained model reaches a prediction accuracy of 75\% or higher.
		
\end{abstract}

%
%

%

%

%
\maketitle


\section{Introduction}

The number of software-related security incidents is constantly increasing. One of the reasons behind this is the large number of vulnerabilities that are reported daily. Compared to the year 2016, the number of identified vulnerabilities doubled in each of the following years \cite{CVEDetails}. In addition, the discovery of vulnerable software modules is a challenging task \cite{wang2019automated}. In comparison to the traditional techniques, the security vulnerability prediction (SVP) approaches are becoming more popular \cite{hovsepyan2012software}. These approaches construct models by utilizing machine learning algorithms in order to identify vulnerable software \cite{chen2019large}. Most existing approaches are based on the identification of vulnerable code sections according to the historical source code data. However, the evolution of methods that the attackers employ as well as the constant updates to the software makes it difficult to identify security issues \cite{jang2014survey}. In addition, software vendors do not always disclose the source code, which makes it challenging to construct datasets for training the model. As a result, further innovative approaches are needed.

Docker containers (DCs) are becoming a highly popular technology that is used for a large variety of purposes \cite{zerouali2018analyzing}. Moreover, many of the containers are publicly available online and can be used for research practices. In addition, they provide a rich data source from which software library-related data can be extracted in order to build high-quality datasets. Coupled with the vulnerability data from the vulnerability databases, this can be used in SVP approaches for the purpose of constructing and training the model. In this paper, a machine learning approach for a cross-analysis between vulnerabilities and software libraries is presented. This is incorporated into the framework called VULNERLIZER, which in comparison to other existing approaches uses the data extracted from Docker images (DIs). It generates a library cohesion graph that includes software library-related information as well as information on how the specific libraries are associated to each other. The approach is automated, extendable, and has a prediction accuracy of 75\% or higher. 

The remainder of this paper is structured as follows: Section 2 examines related work regarding machine learning and matching approaches used for vulnerability analysis. Section 3 discusses the datasets used in the proposed approach. Section 4 presents the VULNERLIZER, as well as its sub-components. Section 5 provides an evaluation of the proposed framework. Finally, Section 6 concludes the paper and provides an outlook on future work. 

\section{Related Work}

In this section, we discuss the related work in the context of applying machine learning approaches as well as the matching approaches for vulnerability analysis.

\subsection{Machine learning approaches}

Harer et al. \cite{harer2018automated} proposed a data-driven approach for automated software vulnerability detection using machine learning. This is achieved by comparing the methods obtained from a source code and build process utilizing multiple machine learning techniques. The results indicated that these techniques can be efficient for predicting the output of static analysis tools at the function level. However, the approach is only applicable to C and C++ programs. Chernis et al. \cite{chernis2018machine} introduced an approach for detecting vulnerabilities by extracting text features from functions in C source code. In this regard, they apply various machine learning classifiers in which they first extract simple features such as the character count, if count and entropy. In addition, they extract complex features such as word n-grams and suffix trees. The results showed a 74\% accuracy rate, which was achieved using character diversity as the baseline requirement. Sabottke et al. \cite{sabottke2015vulnerability} conducted a quantitative and qualitative analysis of vulnerability information found on Twitter. Furthermore, they considered the vulnerability data available in public databases. In order to analyze the collected data, Sabottke et al. applied support vector machine classifiers. As a result, they were able to identify more vulnerabilities exploited in the real-world than the ones for which proof-of-concept is available. This resulted in a lower false-positive detection rate compared to the CVSS-based detectors. Movahedi et al. \cite{movahedi2019cluster} clustered vulnerabilities according to text information available within their own records. In doing so, they simulated a mean value function by relaxing the monotonic intensity function assumption. The approach was applied to multiple operating systems (OSs) and web browsers. The results showed that the clustering approaches perform better and provide more accurate results compared to the non-clustering approaches. Shahzad et al. \cite{shahzad2012large} conducted an exploratory study on the vulnerabilities disclosed from 1988 to 2011. In this regard, they investigated the following seven dimensions: (1) phases in the life cycle of vulnerabilities, (2) evolution of vulnerabilities over the years, (3) functionality of vulnerabilities, (4) access requirements for exploitation of vulnerabilities, (5) risk level of vulnerabilities, (6) software vendors, and (7) software products. The results indicated that the number of reported vulnerabilities did not increase since the year 2008. Moreover, the most exploited vulnerabilities included buffer overflow, DDOS, and privilege escalation. Huang et al. \cite{huang2010text} performed text clustering of vulnerabilities obtained from the National Vulnerability Database (NVD). They applied a cluster overlap index in order to evaluate simple K-means, bisecting K-means, and batchsom clustering algorithms. In addition, they analyzed dimensions such as the vulnerability location, causes, and exploit effects. 

\subsection{Matching approaches}

Umezawa et al. \cite{umezawa2018threat} proposed a threat analysis approach that uses data from vulnerability databases as well as system design information. Initially, they generated attack trees in order to calculate the probability that a safety incident would occur. Furthermore, they generated a set of keywords according to the attack trees and matched them to entries from the vulnerability databases. The results showed that the approach could identify up to 20 vulnerabilities, including the exact matches. With this in mind, Gegick et al. \cite{gegick2005matching} developed a similar approach for matching the attack patterns to vulnerabilities during the design phase. In comparison to the previous approach, they applied regular expressions in order to encapsulate the steps from the attack trees. These are then later matched to vulnerabilities that are extracted from four different databases. The results showed that applying this matching technique can increase security awareness early in the software development lifecycle. Dong et al. \cite{dong2019towards} introduced an approach to detect the inconsistent information between entries from the NVD, unstructured CVE entries, and vulnerability reports. This was achieved by extracting vulnerable software names and versions, and then comparing it to other entries through the application of the ground-truth evaluation. The results showed that the approach is highly accurate and that the presence of inconsistent vulnerable software versions is significant. Pham et al. \cite{pham2010detection} proposed a tool to detect recurring vulnerabilities on systems that reuse the source code or software libraries. The tool applies two different techniques for modeling and matching vulnerable code segments across different systems. The results demonstrated that the approach managed to identify vulnerabilities with high accuracy as well as to report potential issues that received no fixes or patches. With that said, none of the aforementioned approaches considered DCs as a data source. The data that they utilized includes source code files and various vulnerability databases. However, we consider real software library data extracted from different machines that are stored as DIs.

\section{Data sources}

In the following section, we present the data sources employed by the proposed framework. These include the NVD (National Vulnerability Database) and Vulners for CVE data acquisition, and DIs for gathering software library-related data.

\subsection{CVE dataset}
\label{sec:CVEdataset}
In order to determine the most suitable data sources for acquisition of the vulnerability data, we considered the sources described in the taxonomy proposed by Sauerwein et al. \cite{sauerwein2018classification}. As a result, the following two vulnerability data sources were used: NVD \cite{NVD} and Vulners \cite{Vulners}. The NVD is the most commonly used vulnerability database among researchers and practitioners. It is managed by the U.S. government, and it contains security checklist references, security-related software flaws, misconfigurations, product names, and impact metrics. It provides a structured data set that can be accessed via an API. On the other hand, Vulners is a vulnerability assessment platform specializing in the representation of vulnerability data. Similar to the NVD, it provides API access through a Python wrapper. The main difference between these two databases is that Vulners also includes zero-day vulnerabilities that get published as soon as they are reported by the community.

The data was extracted from both of the aforementioned databases using customized data-extraction adapters and stored in the local database. The data consisted of 122.364 CVEs related to 48.833 different software libraries. The CVE dataset was split into two parts: training set and test set. The training dataset was used for populating and training the data model. On the other hand, the test dataset was used to verify the predictions generated by the trained model. More precisely, the exact split consisted of 80\% (103246 CVEs) for the training dataset and 20\% (20649 CVEs)for the test dataset. The 80/20 split was chosen as it provides a high number of data for training the model, thereby allowing it to make quality predictions. Furthermore, it also ensures that there is enough data available that can be used for result verification. The training data was further split into 62.5\% of our training data (50\% of total data) for populating the model, while the remaining 37.5\% (20\% of the total data) was used for adjusting the weights of the model.

\subsection{Docker Images dataset} 

Besides the CVE data, the framework also utilizes software library-related information. This data was gathered from various DIs obtained from the DockerHub \cite{DockerHUB}. The reason behind this is that we were able to quickly gather the large number of data on vulnerable software libraries that are installed on Debian systems. The library is considered vulnerable if it matches the name and version of a library reported in the CVE database. With that said, the images were gathered using specific tags, which were used to search the DockerHub. The extracted information included the name, version, and a brief description of the software. The two tags that were applied are "vulnerable" and "debian". The packet manager that was used for obtaining the library data was only applicable to debian-based systems. As a result, other systems were not considered. The library data obtained from the machines identified using the "vulnerable" tag was used for the population of the model. This allowed the retrieval of a high number of vulnerable libraries. In addition, it also increased the possibility of having multiple vulnerable libraries on the same machine allowing us to better train the model using more data. On the other hand, the library data obtained from the machines identified using the "debian" tag was applied in the evaluation. The exact numbers regarding the DIs dataset can be found in Table ~\ref{tab:DockerImageNumber}.

\begin{table}[!htb]
	\centering
	\small\addtolength{\tabcolsep}{-3pt}
	
	\begin{tabular}{ |c|c|c|c| } 
		\hline
		Tag & Docker img & Img /w library info & Img /w vulns \\
		\hline
		Vulnerable & 116 & 115 & 86 \\
		Debian & 1054 & 360 & 359 \\
		\hline
	\end{tabular}
	
	\caption{Docker images dataset \label{tab:DockerImageNumber}}
\end{table}

\section{Framework}
\label{sec:ToolArchitecture}

\begin{figure*}[ht]
	\centering
	\includegraphics[width=0.85\linewidth]{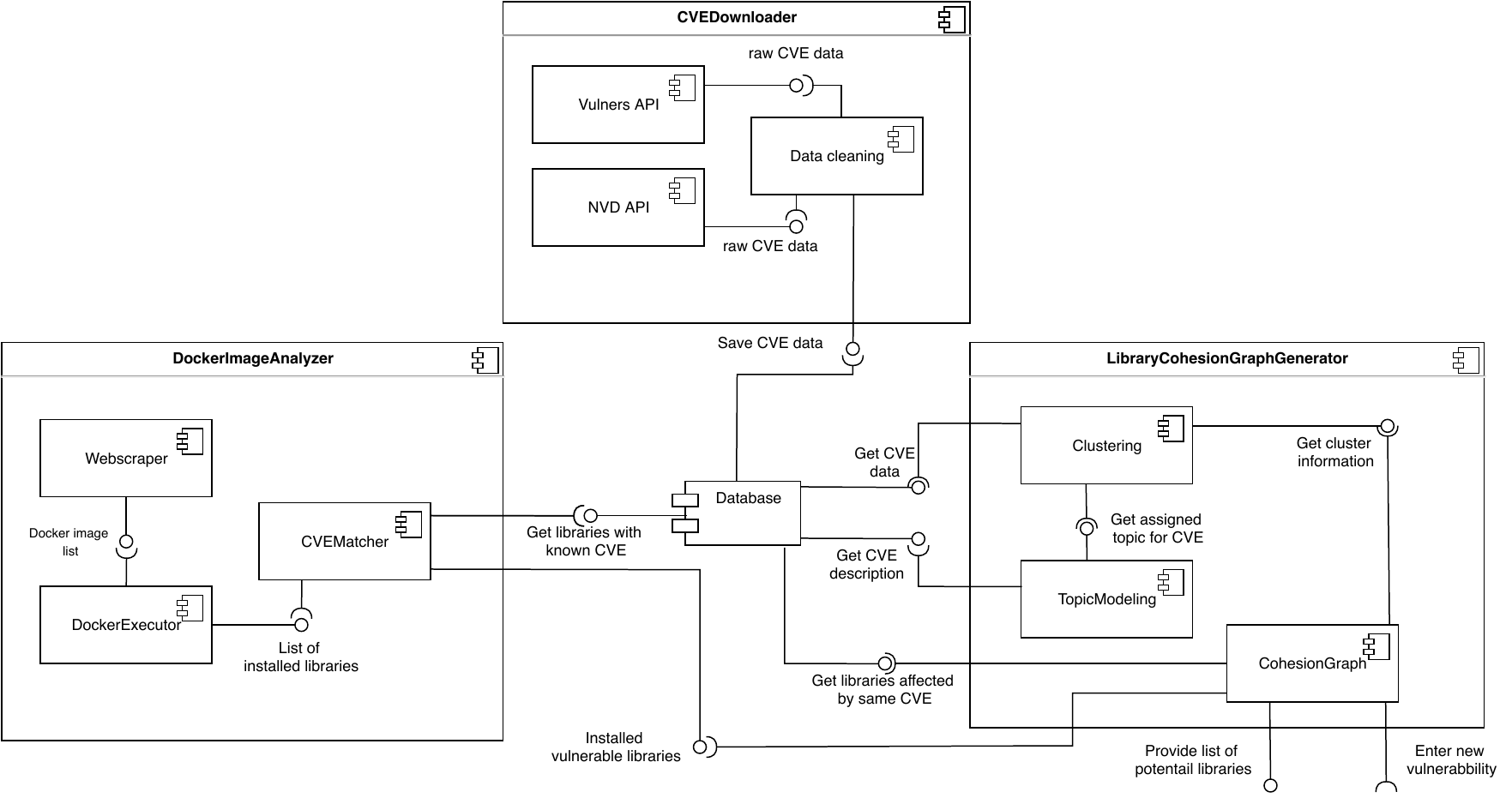}
	\caption{Architecture of the VULNERLIZER}
	\label{fig:workflow}
	\centering
\end{figure*}

The VULNERLIZER is composed of the following three main components that operate together: CVE Downloader, Docker Image Analyzer and Library Cohesion Graph Generator (see Figure \ref{fig:workflow}). In this section, we describe each component in more detail as well as the proposed approach.

\subsection{CVE Downloader}
The CVE Downloader module instantiates a MariaDB database and stores the data obtained from NVD and Vulners (see Section \ref{sec:CVEdataset}). For the extraction of the CVE data from the Vulners, an API wrapper for Python was used. On the other hand, the data from NVD was gathered using a simple http request. As a result, a compressed file that consists of CVE data within a specific time frame was retrieved. This data was then extracted, cleaned, merged, and parsed into the database format.

\subsection{Docker Image Analyzer}

Docker Image Analyzer component conducts an analysis using DIs in order to extract vulnerable libraries.  During the search, the goal was also to gather older images as they might be affected by multiple vulnerabilities, thereby allowing us to build a good training set. We gathered 1250 images in total that were used in the generation of the Library Cohesion Graph. Unfortunately, it was not possible to add more DIs to the set due to the large image processing time. In the following phase, each of the gathered images is analyzed using the AWS EC2 machines.  Once all the machines are started, a FCFS (First Come First Served) approach is applied in order to distribute the images from the previously gathered list to the different machines. The dpkg packet manager is used to get information regarding the installed software libraries, including its name and version. This data is then stored locally for further analysis. Once all the data is acquired, each library is matched with the records from the local database to determine whether it is vulnerable.

\begin{table*}[!htb]
	\centering
	\small
	\setlength\tabcolsep{0.5pt}
	\begin{tabular}{ |c|c|c|c|c|c|c|c|c|c|c| } 
		\hline
		\multicolumn{11}{|c|}{Topics} \\
		\hline
		Topic No.&1&2&3&4&5&6&7&8&9&10 \\
		\hline
		Category&SQL injection&\makecell{Denial of \\service}&\makecell{Cross site \\scripting}& \makecell{Remote execution \\via URL \\parameters}&Other&\makecell{Obtain \\sensitive \\information}&\makecell{Exploits in \\Adobe \\software}
		&\makecell{Vulnerabilities \\on Microsoft \\servers}&\makecell{Remote \\directory \\traversal}&\makecell{Buffer \\overflow} \\
		\hline
		Word 1&sql&service&xss&php&unspecified&servers&earlier&windows&dot&buffer\\
		Word 2&injection&denial&scripting&parameter&vectors&android&acrobat&server&files&overflow\\
		Word 3&commands&cause&cross&inclusion&oracle&the&adobe&microsoft&directory&code\\
		Word 4&execute&memory&site&url&unknown&information&reader&r2&traversal&execute\\
		Word 5&parameter&crash&html&remote&affect&sensitive&exploitation&vulnerability&read&based\\
		\hline
	\end{tabular}
	
	\caption{Most impactful words per topic from the model \label{tab:TopicModel}}
\end{table*}

 
\subsection{Vulnerability Clustering}
\label{sec:Clustering}

In order to gain insights regarding the obtained vulnerabilities and libraries, clustering algorithms were applied. In this regard, we used K-means as a centroid-based algorithm and Density-Based Spatial Clustering of Applications with Noise (DBSCAN) as a density-based algorithm. These were chosen because they are able to cover various cluster sizes. The vulnerability and software library data from the local database is used as an input for clustering. However, it is necessary to parse it into the correct input format by normalizing numerical values and encoding textual values into numerical representations. The component also provides an option to reduce the number of feature dimensions that are used for clustering. This is implemented by applying principle component analysis (PCA) on the feature set of all entries to reduce the dimension to a specified value. This cancels the effect of overfitting the clustering model. The K-means algorithm resulted in 5 to 6 centroids depending on the feature set and permutation being used.

To augment the feature set that can be used for clustering, a natural language processing approach is applied to the descriptions of the CVE dataset. This is done in order to identify various topics in the description of the CVEs, which can be used as an additional input for the clustering algorithms. The approach that was used for topic modeling is a Non-negative Matrix factorization (NMF). Using that approach, it was possible to quickly iterate over the CVE dataset and obtain a large set of different topics. The module is able to parse, clean and tokenize each CVE description. The results showed that by using 10 topics, it was possible to cover a majority of relevant vulnerability groups. Table \ref{tab:TopicModel} demonstrates the results of the topic modeling approach with a set number of 10 topics, wherein the top five most impactful keywords per topic are shown. These were manually assigned and can be seen in the category row.

\subsection{Library Cohesion Graph Generator}
\label{sec:GraphApproach}
The Library Cohesion Graph Generator represents the core component of the VULNERLIZER. It serves to make predictions regarding vulnerable libraries based on the relation to the library that was used as an input.

\subsubsection{Architecture}
\label{sec:ArchitectureGraph}

The architecture of the Library Cohesion Graph is similar to the traditional graph with weighted undirectional edges. Each node stores software library-related information. It contains the following data:
the name of the library, the number of times a given library was affected by a vulnerability from the CVE dataset, the number of times a library was identified as vulnerable on a machine from the Docker Image Analysis (DIA) component, the time between all reported vulnerabilities for a particular library and, clusters of CVEs that affect this library. Each edge stores information regarding the relationship between the two libraries, where each library represents a node, and they are connected with this edge. They contain the following data: the number of times the start and end node libraries are affected by the same CVE, the number of times the two libraries were identified as vulnerable on the same machine in the DIA and the set of weights that are unique for every edge.

The aforementioned weights are used to calculate a cohesion score based on the sum of weighted sub-scores. The weighted sum of the scores is then computed and wrapped by a sigmoid function, which is a standard activation function applied for predicting the probability. The result approximates the possibility of another library being vulnerable based on the input CVE. The sigmoid activation function was used because it allows the use of backpropagation with the gradient descent approach to automatically fine tune the individual weights of each edge. As a result, the accuracy of predictions was increased. The training approach used for training the data model is described in Section \ref{sec:Training}. 


In the following, we provide detailed explanations regarding the equations and notations used to calculate the individual scores. In Equation \ref{eq:cveScore}, the $CVE_{edge}$ represents the number of times a CVE entry was assigned to this particular edge. The $CVE_{start}$ and $CVE_{end}$ was obtained using the same procedure with the difference of employing the start and the end node instead of an edge. The $cve\_score$ was calculated by taking the number of occurrences of both libraries in the same CVE and dividing it by the average number of occurrences of the start and end node in our CVE data set.

\setlength{\abovedisplayskip}{-5pt} \setlength{\abovedisplayshortskip}{-5pt}
\setlength{\belowdisplayskip}{-5pt} \setlength{\belowdisplayshortskip}{-5pt}

\begin{equation}
\label{eq:cveScore}
\resizebox{2in}{!}{$cve\_score = \frac{CVE_{edge}}{(CVE_{start} + CVE_{end}) / 2}$}
\end{equation}

In Equation \ref{eq:machineScore}, the $machine_{edge}$ marks the number of times that the library on a certain edge was identified as vulnerable on a specific machine. We calculated the $machine\_score$ similar to the $cve\_score$ by replacing the CVE occurrences with the machine data we acquired in the DIA.

\begin{equation}
\label{eq:machineScore}
\resizebox{2.4in}{!}{$machine\_score = \frac{machine_{edge}}{(machine_{start} + machine_{end}) / 2}$}
\end{equation}

In Equation \ref{eq:clusterScore}, the $Cl^{node}_{i}$ is the set of clusters assigned to the specific node, where $s$ stands for start node and $e$ for the end node of the current edge at the cluster index $i$. The $cluster\_score$ is computed by calculating the overlap between each cluster between the start and end node, and then dividing it by the total number of cluster assignments. By doing this, it is possible to calculate a score that represents the average overlap between clusters for the start and end node.

\begin{equation}
\label{eq:clusterScore}
\resizebox{3in}{!}{$cluster\_score = \frac{\sum_{i = 0}^{len(Cl^{s} \cup Cl^{e} )} min(Cl^{s}_{i},Cl^{e}_{i}) / max(Cl^{s}_{i},Cl^{e}_{i}) }{len(Cl^{s} \cup Cl^{e} )}$}
\end{equation}

For Equation \ref{eq:clusterMatchScore}, the $cluster\_match\_score$ is calculated by determining how often the assigned cluster of the new vulnerability was assigned to the start and end node. This is then divided by the sum of all assigned clusters. Hence, it is possible to represent the percentage of the cluster for the new vulnerability compared to the total number of assigned clusters for the libraries of the current edge.

\begin{equation}
\label{eq:clusterMatchScore}
\resizebox{2.1in}{!}{$cluster\_match\_score = \frac{Cl_{new\_vul}}{\sum_{i = 0}^{len(Cl)}{Cl^{s}_{i} + Cl^{e}_{i} }}$ }
\end{equation}

In Equation \ref{eq:timeScore}, $time\_score$ is determined by calculating the difference between the average time of vulnerabilities of the target node and the time since the last vulnerability was reported for this node. This is then divided by the highest reported time between the vulnerabilities of the target node. This allows the calculation of the score that determines whether the input vulnerability fits in the historical time data of the target node. With this in mind, $avg\_time^{tar-*-get}$ represents the average time between reported CVEs on the target library and $times\_between\_vulnerbale^{target}$ denotes the list of times between reported CVEs for the target library. 

\begin{equation}
\label{eq:timeScore}
\resizebox{2.6in}{!}{$time\_score = \frac{avg\_time^{target} - time\_since\_last}{max(times\_between\_vulnerable^{target})}$ }
\end{equation}

Equation \ref{eq:sigmoidFunction} calculates the sum of all individual scores with their corresponding weight and uses the sigmoid function as an activation function to determine the activation value of the edge. The higher the activation value is, the stronger the relationship between the two nodes is.

\begin{equation}
\label{eq:sigmoidFunction}
\resizebox{2.8in}{!}{$activation = sigmoid(\sum_{i = 0}^{len(weights)}{weight_{i} * score_{i}})$ }
\end{equation}

\begin{figure}[ht]
	\resizebox{\hsize}{!}{\includegraphics[width=0.2\linewidth]{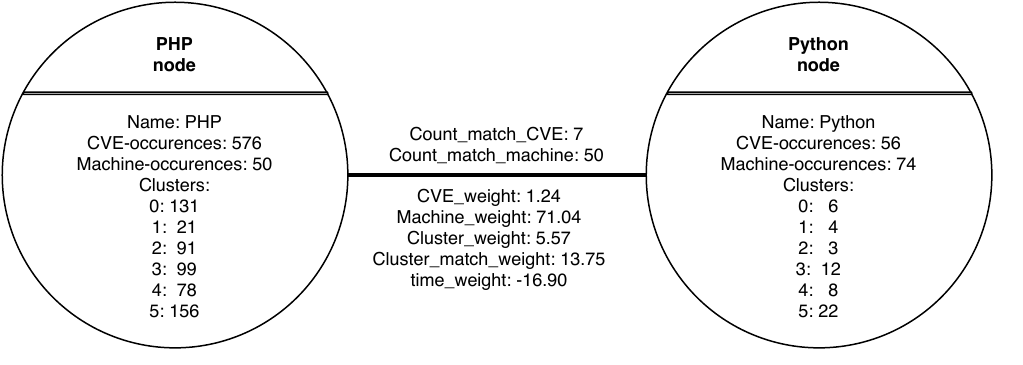}}
	\caption{Example of the Cohesion model with two nodes}
	\label{fig:GraphDetailedNumbers}
	\centering
\end{figure}

The information regarding the new vulnerability, including the assigned cluster and timestamp, are not stored in the model. However, this will be used to predict potential vulnerable libraries based on the input library or the CVE entry. An example of the cohesion model with two nodes is shown in Figure \ref{fig:GraphDetailedNumbers}.

\subsubsection{Population of the model}

The population of the model was conducted using a portion of the training data set. In this regard, we iterate over each CVE entry in the dataset and record whether a certain library was affected by it. In addition, we also check which of the other libraries are also affected by this particular CVE. This is done in order to determine if there is a connection between the two library nodes. The resulting data is then stored in the model by adding the corresponding nodes. If the node already exists, the occurrence count for the library is increased. The same procedure is conducted with the edges. If the combination of the libraries has not been recorded, a new edge between the two corresponding nodes is created. Otherwise, the number of occurrences on the specific edge is increased. An additional method for populating the model involves using the data acquired from the DIA. This data identifies the relations between vulnerable software libraries on the same machine, as well as the occurrence of a library being identified as vulnerable on any machine. With the provided tag, the module will automatically read the data and use it for the population of the model. Finally, the last input for the cohesion graph is the result of the cluster analysis. This is achieved by adding each cluster of CVE entries to the library nodes affected by these CVEs. The information is stored as a dictionary in the library nodes. It indicates how many CVEs from a specific cluster affected a particular software library.

\subsubsection{Prediction of vulnerable libraries}

The main goal of the VULNERLIZER is to conduct an analysis between software libraries and vulnerabilities. In order to achieve this, individual scores were calculated first, as previously mentioned in Section \ref{sec:ArchitectureGraph}. In the following step, a weighted sum of these individual scores was computed. The accumulated score is then fed to an activation function that yields the activation of a given edge between the two libraries in regards to the target library being vulnerable. In order to predict all libraries that might be vulnerable based on a certain library or a CVE entry, we check the library nodes that are affected. Since the input does not have an assigned cluster, we use the clustering algorithm to assign this new entry to a cluster in order to make future predictions. Then, we calculate each activation for all the edges that are connected to the initial library node. Consequently, it is possible to see if the activation exceeds a defined threshold. In case the threshold is lower than the activation of the current edge, the library that is connected to the initial node using that particular edge is added to the list of potentially vulnerable libraries.
Once all the edges connected to the initial node are analyzed, the same analysis is conducted starting from the nodes that are stored in the list of potential libraries. The depth of this look-up approach can be adjusted according to the defined threshold. This is repeated until the given depth is reached or no new potential libraries are added to the final result.
By applying this procedure, we wanted to predict libraries that could potentially be vulnerable because they are related to the library that is used as an input or a CVE entry.

\subsubsection{Training}
\label{sec:Training}

In order to increase the accuracy of predictions of the model, the training of the Library Cohesion Graph was conducted using 37.5\% of the total training data set. This procedure is similar to the one employed by neural networks. In this regard, a gradient descent approach was used to minimize the loss function for the predictions. This is achieved by applying backpropagation, wherein the weights on the individual edges between library nodes are adjusted. The alteration is based on the result of the current prediction. Furthermore, the data in the form of CVE entries with all the supplementary information is used as an input for the clustering algorithms. In addition to the selected features, the date of the publication of the CVE entry is appended as well. This is significant because it allows the calculation of the $time\_score$ from Equation \ref{eq:timeScore}.


A CVE is eligible as an input case for training of the model in case there is at least one other CVE that exists with a publishing date that is later than the publishing date of the input CVE. For this reason, it is possible for the model to identify potential targets. In the first step, the database is scanned for libraries that are affected by the CVE that was published in the next $n$ days after the initial CVE. For example, the $n$ value can be set to zero, which would result in the model being trained to find vulnerabilities that occur on the same day as the input CVE. Another example would be to set the $n$ to a higher value. This would adapt the model to incorporate vulnerabilities that will occur up to $n$ days after the initial input CVE. The found libraries are then stored as target libraries for the initial CVE of this training case. This means that when the initial CVE is used as an input for a prediction, a prediction is considered correct if the predicted library is in the target list of the input CVE. The CVE entries with target lists are then generated for each CVE entry in the training set and stored for the training process. For the training of the model itself, we iterate over the training dataset that contains CVE and the target combinations to generate predictions of potentially vulnerable libraries. These predictions are based on the current weights on the edges between the initial library from the CVE and target libraries. In the following step, the prediction is evaluated with a loss function to determine the prediction loss. This is shown in Equation \ref{eq:lossFunction}. In this regard, we iterate over each entry in the CVE training dataset and repeat the aforementioned procedure for 50 iterations. This is done in order to optimize the weights per edge. The number of iterations can be customized depending on the available resources, such as computing power and time. As a result, each iteration adjusts the individual weights to reduce the loss function and increase the accuracy of the model.


\setlength{\abovedisplayskip}{-5pt} \setlength{\abovedisplayshortskip}{-5pt}
\setlength{\belowdisplayskip}{-5pt} \setlength{\belowdisplayshortskip}{-5pt}

\begin{equation}
\label{eq:lossFunction}
\resizebox{2.3in}{!}{$ loss = 
	\begin{cases}
	1 - prediction, & \text{if target is vulnerable } \\
	1 + prediction,              & \text{otherwise}
	\end{cases} $ }
\end{equation}

\section{Evaluation}
\label{sec:Evaluation}

In the following section, we provide an evaluation of the data model in order to evaluate its effectiveness. We first discuss the method that was applied and its role in verifying the data model. Afterwards, we present the results of the analysis.

\subsection{Method}

The approach used for the evaluation of the model verifies the predictions using the CVE test dataset. The test dataset was prepared using the same methodology that was applied in the preparation of the training dataset. This is done in order to generate a list of target libraries for each CVE entry in the test dataset. Based on the target list, predictions are made for each CVE entry in this set. By doing this, it is possible to determine which libraries could potentially be vulnerable. A library is defined as vulnerable for a specific input case if there exists a CVE related to this library that is published after the CVE that was used as the input for this prediction. In order to cover a wide variety of factors that might influence the accuracy of the predictions, we ran different permutations of factors that could potentially have the highest impact on the predictions. In addition, we compared them in order to identify if some of these factors have a higher influence on the predictions compared to the others.
Consequently, we first defined seven feature sets that were used for the clustering approach in the training and testing of the model. These feature sets contain dimensions from the CVE datatables \footnote{https://bit.ly/vulnerlizer}. Furthermore, it is also stated whether the topic modeling approach was applied for clustering.
The analysis was performed using each feature set for different permutations of the data model. This is done in order to compare them to each other in terms of the achieved accuracy. In the following, we provide a brief description of each permutation:

\noindent \textit{Default} permutation represents the base from which all the other adaptations were made. It uses the full dimension of the feature vector for clustering, wherein the topic modeling was set to work with 10 unique topics. Furthermore, it utilizes only the topic with the highest matching score as an input for clustering. This means there is only one input from the topic modeling that represents one topic assigned to the CVE. Finally, the training of the model was set to run for 50 iterations, while the input tag for populating the model with the libraries from DIs was set to "vulnerable". \noindent \textit{NoTraining} permutation is similar to the \textit{default} permutation. However, it did not use the training part of the data model. This means that all weights are set to their initial value of one. As a result, each sub-score is weighted the same for the cohesion score and the activation of an edge. This permutation was used to verify if the implemented training approach in the data model increases the accuracy of the model's predictions. \noindent \textit{20Topics} permutation utilizes a different number of topics for the topic modeling approach. Instead of using 10 topics, 20 topics were applied. This investigates whether the number of used topics has a significant impact on the accuracy of predictions. \noindent \textit{Multiple\_topics} permutation uses topic modeling as a feature for clustering in a different fashion. Instead of using a single topic with the highest correlation value, the correlation value for each topic was applied. This means that the number of dimensions of the feature vector for clustering was increased by nine. Using that permutation, it is possible to test whether the topic modeling performs better with a single assigned topic per vulnerability compared to the correlation values for all the topics. \noindent \textit{1dimension} permutation uses the dimension reduction feature of the VULNERLIZER. This reduces the number of dimensions of the feature vector used for clustering to one dimension. As a result, it is possible to check if the \textit{default} permutation created issues, such as overfitting in the clustering of vulnerabilities. \noindent \textit{Lessepochs} permutation uses the lower number of iterations for the training of the model. Instead of 50, 20 iterations were used. This verifies whether the number of iterations used for training the model has an impact on the final accuracy or not. \textit{Debian} permutation uses the "debian" tag instead of the "vulnerable" tag for the population of the graph in regards to the software library data obtained from DIs. This determines if the source data from the DIA has any impact on the accuracy.

\subsection{Evaluation based on CVEs}

In order to further prepare the data for evaluation, it was necessary to iterate over all CVE entries in the dataset and assign each entry an ID for one of the clusters from the training process. Moreover, a target library list was assigned to each CVE entry by looking at subsequent CVEs of the input CVE. This was done in the same way as the preparation of the training dataset. In this case, we considered a library as a target if it was affected by a CVE within 45 days of publication of the initial input CVE. The 45 days were chosen since this time period was considered broad enough to evaluate the performance of the model in regards to predicting the libraries that could become vulnerable after the report of a new CVE. Afterwards, this data was used for model verification by feeding each initial CVE into the model and comparing the generated list of potentially vulnerable libraries with the target list of the CVE. We rated a prediction as correct if the predicted library was part of the target list and as false if the prediction was not part of the list. The average accuracy ($\frac{correct \, predictions}{all \, predictions}$) was built using all CVEs in the test run. Using that approach, it is possible to verify the accuracy of the generated predictions. However, it is not possible to strictly assign a relation between the CVEs. For this reason, there is a chance that not all of the related CVEs are predicted. Nevertheless, the accuracy was more important than predicting every possible library in regards to the libraries that might be potentially vulnerable. 

\begin{figure}[ht]
	\centering
	\resizebox{\hsize}{!}{\includegraphics[width=0.6\textwidth]{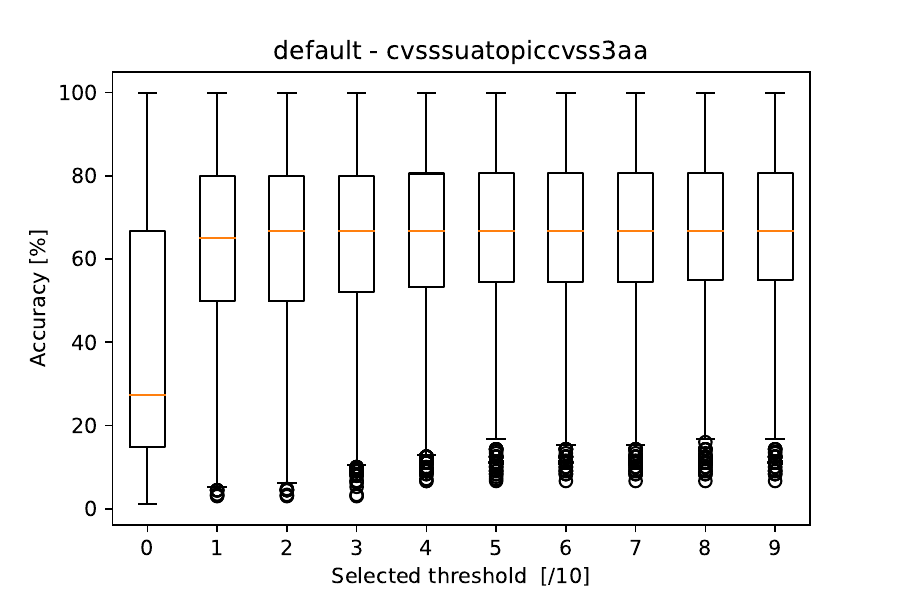}}
	\caption{Trend of accuracy based on threshold}
	\label{fig:accTrend}
	\centering
\end{figure}

Furthermore, if the verification data does not contain a library represented by a node in the model, it would be dropped from the evaluation. In this case, it is possible that a newly reported vulnerability was related to a library that was not found during the analysis. Therefore, we are not able to make predictions for these specific test cases. As a result, they are omitted from the evaluation. This reduced the number of possible test cases per permutation to an average of 19900 test cases. In all permutation / feature\_set combinations, except for the permutation without the training, the accuracy trend based on the given threshold followed the same pattern. The accuracy values for predictions with a threshold of 0 were the lowest for the whole dataset, whereby all edges were used for predictions. On the other hand, the accuracy values were higher for a threshold of 0.1 or higher. While the initial jump in the accuracy from a threshold of 0 to 0.1 was the highest with an average increase by 24.5\%, further increases in the threshold did not yield larger differences in the accuracy values. The default permutation is shown in Table \ref{fig:accTrend}, in which the feature\_set that resulted in the highest average accuracy was used.

In regards to all permutations, the threshold of 0.9 had the highest prediction accuracy. This was expected because the higher threshold reduces the number of activated edges, which yields a lower number of false predictions for the input library. However, most of the correct predictions are retained because the weights were adjusted due to the training of the model. As a result, a higher activation on historically correct predictions was generated. 

\subsubsection{Evaluation using different permutations}

\begin{figure}[ht]
	\centering
	\resizebox{\hsize}{!}{\includegraphics[width=0.6\textwidth]{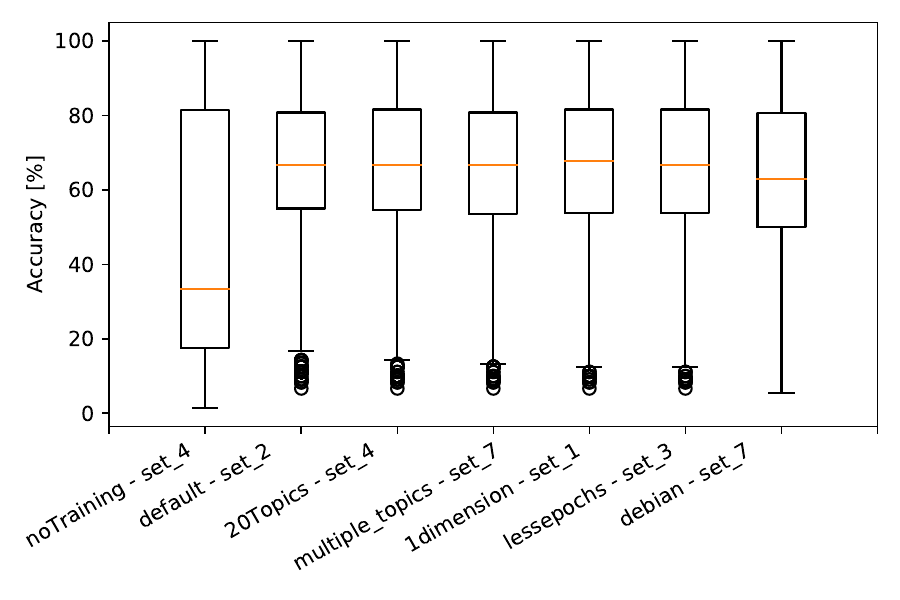}}
	\caption{Accuracies for permutations with best performing feature set $<$permutation$>$ - $<$feature\_set$>$}
	\label{fig:comparison}
	\centering
\end{figure}

Figure \ref{fig:comparison} displays the comparison between accuracies over the different permutations. Each boxplot displays all accuracies for 19900 test cases for the given permutation. The threshold of 0.9 was chosen each time as it had the best resulting accuracy for all permutations and feature set combinations. In addition, the feature set that performed the best was selected for the given permutation. The x axis of the graph shows the combination $<$permutation$>$ - $<$feature\_set$>$. It can be seen that compared to the noTraining and the default permutation, the training of the model increased the median accuracy of the predictions by 21.68\% (from 45.4\% to 67.1\%). The other permutations, except for the debian permutation, resulted in similar median accuracy values. However, in the non-default permutations, the distribution of the accuracy values is shifted slightly to the lower ranges. Similarly, the debian permutation resulted in lower accuracy for all the data points. On the other hand, the default permutation was the best performing one. Generally, the changes in permutations slightly decreased the accuracy of predictions. This can be seen from the data displayed in Figure \ref{fig:comparison}, wherein the median accuracy remains the same. On the other hand, there are more data points in the lower accuracy range for the other permutations. This proves that the values chosen for the default permutation are better than the altered ones used in the other permutations. According to the results, most of the data is located around the mean accuracy. However, each permutation has some outliers that are lower than the minimum accuracy in the respective boxplot. Therefore, we analyzed the accuracy with regards to the number of predicted vulnerable libraries per test CVE. For this purpose, we build the average accuracy values of all test cases with the same number of predicted libraries. The purpose of this is to see the accuracy distribution based on the number of predicted libraries per new CVE. This indicates whether the model is able to predict a high number of libraries with confidence. We used this approach to check if we can identify the outliers below the minimum value, as shown in the boxplots in Figure \ref{fig:comparison}.

\subsubsection{Evaluation using the best feature set} 

\begin{figure}[ht!]
	\centering
	\includegraphics[trim={2.5cm 6cm 2.5cm 4cm},clip, width=\linewidth]{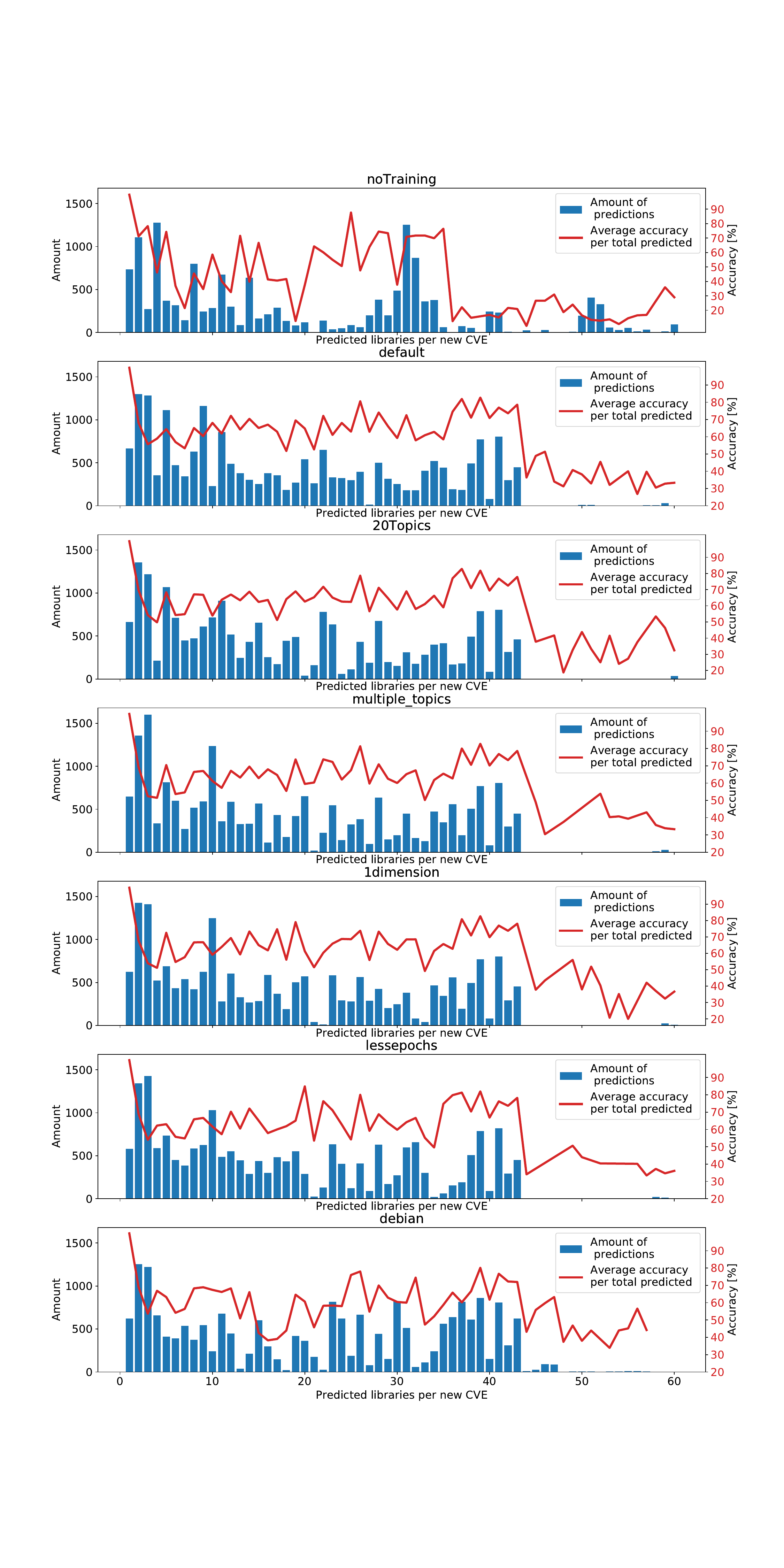}
	\caption{Accuracy and number of predictions based on correctly predicted libraries}
	\label{fig:accperfound}
	\centering
\end{figure}

Figure \ref{fig:accperfound} demonstrates the number of predictions made per CVE per permutation using the best set. The best set is defined as the feature set where the highest accuracy values are achieved per permutation. The red line indicates the average accuracy for the number of predicted libraries per CVE. We decided to cap the range on the x axis at 60 because only a small number of permutations had more than 60 test cases. These did not often occur though, which is why we consider them as outliers and ignore them in order to enhance the readability of the graphs.
According to the chart, the model had a good performance when predicting up to 43 libraries per input CVE. From 43 predicted libraries onwards, the accuracy of predictions drops together with the number of predictions due to the outliers. This is true for each permutation except for the \textit{NoTraining} permutation. In this case, the average accuracy of predictions has multiple drops in relation to the total number of predictions (y-axis) per CVE. Furthermore, these drops appear earlier compared to the other permutations. Another observation is that there are more peaks for the specific numbers of predictions per CVE. In relation to other permutations, the distribution of the occurrences of specific numbers of predictions per CVE is distributed better over the displayed range. This is probably due to the missing adjustments of weights for these permutations. This indicates that the trained model is able to find more potentially vulnerable libraries compared to the model without training. The training of the model allows reinforcing connections between libraries that are related and weakening connections for non-related libraries.

An example of a trained permutation is the \textit{Debian} permutation, where the average accuracy results range from 14 to 19 predictions per initial input. This is lower compared to the other permutation sets. Moreover, the number of test cases in which the number of predictions was between 13 and 19 per input CVE is also lower in the \textit{Debian} permutation. This is probably due to the lower average number of identified vulnerable libraries on the machines with the "debian" tag (see Table \ref{tab:DockerImageNumber}) compared to the machines with the "vulnerable" tag from the DIA. In addition, this permutation resulted in a shift of the number of correctly predicted libraries to the right side. This means that a higher number of libraries is predicted more often. Due to the lower accuracy in the \textit{Debian} permutation for the test cases with the lower number of predictions, the overall accuracy for this permutation is noticeably lower as one can see in Figure \ref{fig:comparison}. This is due to the lower number of identified vulnerabilities per machine in the DIA. Furthermore, the average accuracy when the number of predictions from the model is low varies stronger between each permutation. This is not the case when the model is predicting more libraries. As a result, the model performs best when predicting multiple libraries (see Figure \ref{fig:accperfound}). This is due to the initial node being well connected within the model. Therefore, the accuracy is higher because more information is available. Having a high number of test cases where the number of predictions per CVE is low, can be justified by the fact that some entry nodes might not be fully connected within the graph. This is the case because the data that was used for the population and training was not linked to any node due to the missing relations obtained from analyzing the datasets. This issue could be addressed by collecting additional DI data and extracting additional software libraries.

\subsection{Threats to validity}

Throughout the development and evaluation of the VULNERLIZER, we considered possible threats to validity. The approach in which the data is extracted from the DIs could be subjective. The packet manager that was used is intended to work only for OSs that are debian-based. As a result, the other OS types were not considered. Furthermore, the image search was conducted using only two different tags. However, we were still able to obtain a high number of data for training and testing of the model. Moreover, the obtained CVE data might not always be up-to-date. This means that the data is as actual as its source. As a result, there might be relevant vulnerabilities that are published after the data gathering is completed. In order to mitigate this issue, the tool updates the database and checks if there are new vulnerabilities each time it is run. Finally, the quality of the descriptions in the CVE data that is used for the topic modeling might be questionable. This relates especially to the newest entries, as they might not have the full description yet. Furthermore, it is worth noting that the descriptions do not need to follow fixed guidelines, and they are written by many different security experts. Thus, CVE descriptions can range from very detailed ones to the ones that are empty. In order to minimize this threat, we considered multiple data sources. In addition, the CVEs with an empty description have a unique topic assigned to them.

\section{Conclusion}

In this paper, we presented a novel approach for cross-analysis between vulnerabilities and software libraries. The approach uses CVE and software library data in combination with clustering algorithms to generate links between vulnerabilities and libraries. In addition, the training of the model is conducted in order to reevaluate the generated links. Finally, the predictions are made using the CVE data from the test set. The results indicate that the VULNERLIZER has the potential to predict future vulnerable libraries based on an initial CVE entry or a software library. The trained model performed best in cases when a high number of predictions was made, which resulted in an accuracy of 75\% or higher. The overall approach could be integrated into secure life cycle of IT companies in order to check if some of the dependencies they use are vulnerable. Future work involves the consideration of other data sources, such as the network data and the data from Git repositories. Furthermore, we plan to apply additional data training methods to improve the model further, as well as to apply this approach to different OSs. 

\paragraph{Acknowledgement} This work was partially supported by the Austrian Science Fund (FWF): I 4701-N.

\bibliographystyle{ieeetr}
\bibliography{main}

\begin{thebibliography}{10}

\bibitem{CVEDetails}
``{CVE} {Details}.'' \url{https://www.cvedetails.com/browse-by-date.php}, 2020.
\newblock Accessed: 2020-04-12.

\bibitem{wang2019automated}
Z.~Wang, Y.~Zhang, Z.~Tian, Q.~Ruan, T.~Liu, H.~Wang, Z.~Liu, J.~Lin, B.~Fang,
  and W.~Shi, ``Automated vulnerability discovery and exploitation in the
  internet of things,'' {\em Sensors}, vol.~19, no.~15, p.~3362, 2019.

\bibitem{hovsepyan2012software}
A.~Hovsepyan, R.~Scandariato, W.~Joosen, and J.~Walden, ``Software
  vulnerability prediction using text analysis techniques,'' in {\em
  Proceedings of the 4th international workshop on Security measurements and
  metrics}, pp.~7--10, 2012.

\bibitem{chen2019large}
X.~Chen, Y.~Zhao, Z.~Cui, G.~Meng, Y.~Liu, and Z.~Wang, ``Large-scale empirical
  studies on effort-aware security vulnerability prediction methods,'' {\em
  IEEE Transactions on Reliability}, 2019.

\bibitem{jang2014survey}
J.~Jang-Jaccard and S.~Nepal, ``A survey of emerging threats in
  cybersecurity,'' {\em Journal of Computer and System Sciences}, vol.~80,
  no.~5, pp.~973--993, 2014.

\bibitem{zerouali2018analyzing}
A.~Zerouali, ``Analyzing technical lag in docker images.,'' in {\em BENEVOL},
  pp.~11--15, 2018.

\bibitem{harer2018automated}
J.~A. Harer, L.~Y. Kim, R.~L. Russell, O.~Ozdemir, L.~R. Kosta, A.~Rangamani,
  L.~H. Hamilton, G.~I. Centeno, J.~R. Key, P.~M. Ellingwood, {\em et~al.},
  ``Automated software vulnerability detection with machine learning,'' {\em
  arXiv preprint arXiv:1803.04497}, 2018.

\bibitem{chernis2018machine}
B.~Chernis and R.~Verma, ``Machine learning methods for software vulnerability
  detection,'' in {\em Proceedings of the Fourth ACM International Workshop on
  Security and Privacy Analytics}, pp.~31--39, ACM, 2018.

\bibitem{sabottke2015vulnerability}
C.~Sabottke, O.~Suciu, and T.~Dumitraș, ``Vulnerability disclosure in the age
  of social media: exploiting twitter for predicting real-world exploits,'' in
  {\em 24th $\{$USENIX$\}$ Security Symposium ($\{$USENIX$\}$ Security 15)},
  pp.~1041--1056, 2015.

\bibitem{movahedi2019cluster}
Y.~Movahedi, M.~Cukier, A.~Andongabo, and I.~Gashi, ``Cluster-based
  vulnerability assessment of operating systems and web browsers,'' {\em
  Computing}, vol.~101, no.~2, pp.~139--160, 2019.

\bibitem{shahzad2012large}
M.~Shahzad, M.~Z. Shafiq, and A.~X. Liu, ``A large scale exploratory analysis
  of software vulnerability life cycles,'' in {\em 2012 34th International
  Conference on Software Engineering (ICSE)}, pp.~771--781, IEEE, 2012.

\bibitem{huang2010text}
S.~Huang, H.~Tang, M.~Zhang, and J.~Tian, ``Text clustering on national
  vulnerability database,'' in {\em 2010 Second international conference on
  computer engineering and applications}, vol.~2, pp.~295--299, IEEE, 2010.

\bibitem{umezawa2018threat}
K.~Umezawa, Y.~Mishina, S.~Wohlgemuth, and K.~Takaragi, ``Threat analysis using
  vulnerability databases--matching attack cases to vulnerability database by
  topic model analysis--,'' in {\em The Third International Conference on
  Cyber-Technologies and Cyber-Systems (CYBER 2018)}, 2018.

\bibitem{gegick2005matching}
M.~Gegick and L.~Williams, ``Matching attack patterns to security
  vulnerabilities in software-intensive system designs,'' in {\em Proceedings
  of the 2005 workshop on Software engineering for secure systems—building
  trustworthy applications}, pp.~1--7, 2005.

\bibitem{dong2019towards}
Y.~Dong, W.~Guo, Y.~Chen, X.~Xing, Y.~Zhang, and G.~Wang, ``Towards the
  detection of inconsistencies in public security vulnerability reports,'' in
  {\em 28th $\{$USENIX$\}$ Security Symposium ($\{$USENIX$\}$ Security 19)},
  pp.~869--885, 2019.

\bibitem{pham2010detection}
N.~H. Pham, T.~T. Nguyen, H.~A. Nguyen, and T.~N. Nguyen, ``Detection of
  recurring software vulnerabilities,'' in {\em Proceedings of the IEEE/ACM
  international conference on Automated software engineering}, pp.~447--456,
  2010.

\bibitem{sauerwein2018classification}
C.~Sauerwein, I.~Pekaric, M.~Felderer, and R.~Breu, ``An analysis and
  classification of public information security data sources used in research
  and practice,'' {\em Computers \& Security}, vol.~82, 12 2018.

\bibitem{NVD}
``National {Vulnerability} {Database}.'' \url{https://nvd.nist.gov/}, 2020.
\newblock Accessed: 2020-02-08.

\bibitem{Vulners}
``Vulners.'' \url{https://github.com/vulnersCom/api}, 2020.
\newblock Accessed: 2020-02-03.

\bibitem{DockerHUB}
``Docker {Hub}.'' \url{https://hub.docker.com/}, 2020.
\newblock Accessed: 2020-02-01.

\end{thebibliography}

\end{document}